\documentclass[aip,jcp,reprint,nofootinbib]{revtex4-2}
\usepackage{graphicx,amssymb,amsmath,mhchem}
\usepackage{hyperref}
\usepackage{color}
\usepackage[dvipsnames]{xcolor}
\begin{document}
\title{Diffuse-Layer Capacitance at the Potential of Zero Charge in Binary Mixtures} 
\author{Yuki Uematsu}
\email{uematsu@phys.kyutech.ac.jp}
\affiliation{Department of Physics and Information Technology, Kyushu Institute of Technology, Iizuka 820-8502, Japan}
\date{\today}

\begin{abstract}
The capacitance of the electric double layer has potential applications in supercapacitors, and theoretical investigations of the double-layer capacitance in binary mixtures are important.  
In this work, we develop the theory of the electric double layer in binary mixtures, and the diffuse-layer capacitance at the potential of zero charge is obtained analytically.
Furthermore, we observe a divergence of the capacitance in the phase diagram, suggesting a surface instability. 
The obtained capacitance is different from that derived using single-liquid approximation unless the preferential solvation energies of cations and anions are the same. 
When the system is close to the surface instability line, the capacitance strongly deviates from the results of single-liquid approximation.

\end{abstract}

\maketitle

\section{Introduction}

The structure of the electric double layer has been studied for a long time because charged solid-liquid interfaces are ubiquitous in nature and industrial applications \cite{Uematsu_2018_1,Uematsu_2021,Becker_2023}.
The Gouy-Chapman model is the simplest and most standard model for the structure of electric double layer \cite{Gouy_1910,Chapman_1913}. 
It considers the thermal diffusion of ions as well as electrostatics, and the electric double layer has a thickness known as a Debye length, $\kappa^{-1}=\sqrt{\varepsilon\varepsilon_0k_\mathrm{B}T/2e^2c_\mathrm{b}}$, where $\varepsilon$ is the dielectric constant of the liquid, $\varepsilon_0$ is the electric permittivity of vacuum, $k_\mathrm{B}$ is the Boltzmann constant, and $T$ is the temperature, $e$ is the elementary charge, and $c_\mathrm{b}$ is the bulk salt concentration. 
The resultant differential capacitance is $C_\mathrm{GC} = \varepsilon\varepsilon_0\kappa \cosh(e\psi_0/2k_\mathrm{B}T)$ where $\psi_0$ is the surface potential. 
At the potential of zero charge (PZC) in the model ($\psi_0=0$), the capacitance is $\varepsilon\varepsilon_0\kappa$, which is proportional to the square root of the salt concentration. 
However, for example, in the case of a mercury electrode at high salt concentration \cite{Grahame_1947}, the dependence of the capacitance at PZC on the salt concentration is weaker than $\sqrt{c_\mathrm{b}}$. 
This discrepancy is partially resolved by introducing the Stern layer at the interface between the solid surface and the diffuse layer.
The differential capacitance is then the direct series of the Stern-layer capacitance, $C_\mathrm{S}$, and the diffuse-layer capacitance, given by $C=(1/C_\mathrm{S}+1/C_\mathrm{GC})^{-1}$. 
Furthermore, the adsorption of counterions onto the interface of the Stern layer improves the shape of the differential capacitance \cite{Grahame_1947}.

When the solvent is altered by binary mixtures, the structure of the electric double layer is affected by two unique properties of the binary mixture.  
The first one is the composition dependence of the dielectric constant of the solvent, and the second one is the preferential solvation of ions by each solvent \cite{Onuki_2004,Onuki_2006,Ben_Yaakov_2009,Onuki_2011}.  
In the electric double layer of the binary mixture, the inhomogeneity of the solvent composition can be coupled with the position-dependent dielectric constant, as well as the preferential solvation of ions.  
There have been some experimental studies on the capacitance measurements in binary mixtures \cite{Aoki_2018, Iwasaki_2023}.
Aoki et al.~ studied the double layer capacitance of platinum electrodes in water-acetonitrile mixtures.  
They observed the capacitance variation only at a water mole fraction in the range of $0$ to $0.2$, and otherwise the capacitance in the binary mixture is the same as that in pure water.  
Iwasaki et al.~ measured the double layer capacitance of aluminum electrodes in water-1-propanol mixtures, and a similar trend was observed. 
Although these results suggest that water preferentially adsorbs to the electrode surface, the reason is still not clear.
The difficulties in experimental measurement of double layer capacitance are that a solid electrodes has surface roughness and behaves like a constant phase element.
From the theoretical viewpoint, although a few preliminary studies on the double layer structure in binary mixtures have been conducted before \cite{Ben_Yaakov_2009, Pousaneh_2012, Pousaneh_2014, Yabunaka_2017}, there is no theoretical study focusing on the capacitance in a binary mixture. 

In this paper, we calculate the capacitance at the PZC in a binary mixture. 
Using a similar model as earlier studies \cite{Ben_Yaakov_2009,Yabunaka_2017}, we analytically calculate the diffuse layer capacitance at the PZC. 
The model accounts for the effect of preferential solvation of ions, as well as the composition dependence of the dielectric constant. 
We derive an analytical expression for the capacitance and the screening length at the PZC in a binary mixture and verify it using a numerical solution. 
We hope that this study helps improve the fundamental understanding of the structure of electric double layers in a binary mixture.
 
\section{Model}
The model described in this section was similar to that in Ref.~\citenum{Ben_Yaakov_2009} (without the square gradient term) or Ref.~\citenum{Yabunaka_2017} (with the square gradient term). 
The Helmholtz free energy of an electrolyte solution in a binary mixture is given by
\begin{equation}
F  = \int  \left[f+f_\mathrm{g}+f_\mathrm{e} \right] d\boldsymbol{r},
\end{equation}
where $f$ is the chemical part of the Helmholtz free energy density, $f_\mathrm{g}$ is the gradient free energy, and $f_\mathrm{e}$ is the electrostatic energy density. 
We assume an isothermal and isochoric process. 
Furthermore, we consider the situation where the volume fraction of the ions is negligible. 
Under these conditions, the Helmholtz free energy density is a function of the local volume fraction of the second solvent $\phi$, the cation and anion concentrations, $c_+$ and $c_-$. 
Assuming the first and second solvents have equal molecular volume, $v_0$, the Helmholtz free energy density at the mean-field level is
\begin{equation}
\begin{split}
& f(\phi,c_+,c_-)  = k_\mathrm{B}T\sum_{i=\pm}c_i\left[\ln(c_i v_0)-1+\alpha_i\phi\right]\\
& + \frac{k_\mathrm{B}T}{v_0}\left[\phi\ln\phi+(1-\phi)\ln(1-\phi)+\chi\phi(1-\phi)\right], 
\end{split}
\label{eq:2}
\end{equation}
where $k_\mathrm{B}$ is the Boltzmann constant, $T$ is the temperature, $\alpha_i$ are the parameters for preferential solvation of ions, and $\chi$ is the parameter for the interaction between two different solvent molecules.  
The gradient free energy is given by \cite{Yabunaka_2017}
\begin{equation}
f_\mathrm{g} = \frac{K}{2}(\nabla\phi)^2,
\end{equation}
where $K$ is the constant with the order of $k_\mathrm{B}T/{v_0}^{1/3}$.
The electrostatic energy density is given by
\begin{equation}
f_\mathrm{e}=\frac{\varepsilon(\phi)\varepsilon_0}{2}(\nabla\psi)^2,
\end{equation}
where $\varepsilon(\phi)$ is the dielectric constant, $\varepsilon_0$ is the electric permittivity of vacuum, and $\psi$ is the local electrostatic potential.  
The dependence of the dielectric constant on the volume fraction of the second component is considered to be linear, as given by
\begin{equation}
\varepsilon(\phi) = \varepsilon_1+(\varepsilon_2-\varepsilon_1)\phi,
\end{equation}
where $\varepsilon_1$ and $\varepsilon_2$ are the dielectric constants of the first and the second solvents.
We assume that the first solvent has a high dielectric constant like water, whereas the second solvent is polar but has a lower dielectric constant. 
A more accurate description for such polar solvents within the free energy formalism is known as the dipolar Poisson-Boltzmann equation \cite{Abrashkin_2007}.
However, we use a simpler model in this work.

Here, we specify the geometry of the system. 
Since we investigate an electric double layer near a planar surface, the coordinate of the system is set as $z \in [0,\infty)$. 
Because the Helmholtz free energy should depend only on extensive quantities (or their densities) and temperature, it is useful to write the electrostatic energy per area in terms of $\phi$, $c_+$, $c_-$.
Defining $F_\mathrm{e}=\int f_\mathrm{e} \,d\boldsymbol{r}$, it is given by
\begin{equation}
\frac{F_\mathrm{e}}{A}= \int_0^\infty \left[-\frac{\varepsilon(\phi)\varepsilon_0}{2}\left(\frac{d\psi}{dz}\right)^2 + \rho\psi \right] dz + \sigma_0\psi_0,
\label{eq:electrostatics}
\end{equation}
where $A$ is the area of the solid surface, $\rho = e(c_+-c_-)$ is the charge density, $e$ is the elementary charge, $\sigma_0 = -\varepsilon(\phi)\varepsilon_0 d\psi/dz|_{z=0}$ is the surface charge density of the solid surface, and $\psi_0=\psi(0)$ is the potential at the surface.
Eq.~\eqref{eq:electrostatics} can be derived from the Poisson equation given by
\begin{equation}
\frac{d}{dz}\left(\varepsilon(\phi)\varepsilon_0\frac{d\psi}{dz}\right) = -\rho.\label{eq:Poisson}
\end{equation}
The derivative form of Eq.~\eqref{eq:electrostatics} with respect to $\delta \phi(z)$, $\delta \rho(z)$, and $d\sigma_0$ is 
\begin{equation}
d\left(\frac{F_\mathrm{e}}{A}\right) = \int^\infty_0 \left[ -\frac{\varepsilon_0}{2}\frac{\partial\varepsilon}{\partial \phi}\left(\frac{d\psi}{dz}\right)^2\delta\phi+\psi\delta\rho\right]dz +\psi_0 d\sigma_0.
\end{equation}
Another advantage of Eq.~\eqref{eq:electrostatics} is that it allows us to derive the Poisson equation and the boundary condition by differentiation of $\psi(z)$ and $\psi_0$. 
Considering the small increment of $\psi(z)\to\psi(z)+\delta\psi(z)$ and $\psi_0\to\psi_0+d\psi_0$ in Eq.~\eqref{eq:electrostatics},
\begin{equation}
\begin{split}
 d\left(\frac{F_\mathrm{e}}{A}\right) = & \int_0^\infty \left[-\varepsilon(\phi)\varepsilon_0\frac{d\psi}{dz} \frac{d\delta\psi}{dz}+ \rho \delta \psi \right] dz + \sigma_0 d\psi_0,\\
=& \int_0^\infty \left[\frac{d}{dz}\left(\varepsilon(\phi)\varepsilon_0\frac{d\psi}{dz}\right)+ \rho\right] \delta\psi dz \\
&+\left(\varepsilon(\phi_0)\varepsilon_0\left.\frac{d\psi}{dz}\right|_{z=0}+\sigma_0\right)\delta\psi_0 = 0,
\end{split}
\end{equation}
yielding the Poisson equation and its boundary condition. 
To derive the boundary condition for $\phi$, we minimize the free energy
\begin{equation}
\frac{F_\mathrm{g}}{A} = \int_0^\infty f_\mathrm{g}dz + h\phi_0,
\end{equation}
where $h$ is the parameter for the interaction between the wall and the solvents.
The minimization of the free energy yields
\begin{equation}
d\left(\frac{F_\mathrm{g}}{A}\right) =  -\int^\infty_0 K\frac{d^2\phi}{dz^2}\delta\phi dz+\left[-K\left.\frac{d\phi}{dz}\right|_{z=0}+h\right] \delta\phi_0,
\end{equation}
where the first term is the additional contribution to the chemical potential for the volume fraction, whereas the second term yields the boundary condition for $\phi$, as given by $d\phi/dz|_0=h/K$.
In order to maintain the homogeneous state ($\phi(z)=\phi_\mathrm{b}$ and $\psi(z)=0$) as the PZC, we set $h=0$ throughout the paper.

To obtain the equilibrium profiles of $\phi(z)$, $c_i(z)$, and $\psi(z)$ for given bulk concentrations $c_\mathrm{b}$ and $\phi_\mathrm{b}$, and the surface charge density $\sigma_0$, we convert the Helmholtz free energy to the grand potential $\Omega$ by Legendre transform as
\begin{equation}
\frac{\Omega}{A} = \frac{F}{A} - \int^\infty_0 \left[\mu_+c_+(z)+\mu_-c_-(z)+\mu_\phi\phi(z) \right] dz, 
\end{equation}
where $\mu_\pm$ are the chemical potentials of the cation and anion, and $\mu_\phi$ is the chemical potential for the volume fraction.
Therefore, the equilibrium conditions are given by 
\begin{eqnarray}
\frac{\delta\Omega}{\delta c_i(z)} &=& 0,\label{eq:chemion} \\
\frac{\delta\Omega}{\delta\phi(z)} &=& 0,\label{eq:chemsol} \\
\frac{\delta\Omega}{\delta\psi(z)} &=& 0,\label{eq:eqpb} \\
\frac{\partial\Omega}{\partial\psi_0} &=& 0, \label{eq:bc} \\
\frac{\partial\Omega}{\partial\phi_0} &=& 0, \label{eq:bc2}
\end{eqnarray}
where Eqs.~\eqref{eq:chemion} and \eqref{eq:chemsol} are the homogeneous conditions for the chemical potentials of the solvents and the ions, and Eq.~\ref{eq:eqpb} is the Poisson equation, and Eq.~\ref{eq:bc} is the boundary condition for the Poisson equation.  
Eq.~\eqref{eq:chemion} yields the concentrations of cations and anions as
\begin{eqnarray}
c_+(z) & = & c_\mathrm{b}\mathrm{e}^{-\Psi(z)-\alpha_+(\phi(z)-\phi_\mathrm{b})},\label{eq:boltzmann1}\\ 
c_-(z) & = & c_\mathrm{b}\mathrm{e}^{ \Psi(z)-\alpha_-(\phi(z)-\phi_\mathrm{b})},\label{eq:boltzmann2}
\end{eqnarray}
where $\Psi= e\psi/k_\mathrm{B}T$ is the dimensionless potential. 
Eq.~\eqref{eq:Poisson} with eqs.~\eqref{eq:boltzmann1} and \eqref{eq:boltzmann2} yields the Poisson-Boltzmann equation in a binary mixture, as given by
\begin{equation}
\begin{split}
& \frac{d}{dz}\left[\varepsilon(\phi)\varepsilon_0\frac{d\psi}{dz}\right] \\
& = -ec_\mathrm{b}\left[\mathrm{e}^{-\Psi(z)-\alpha_+(\phi(z)-\phi_\mathrm{b})}-\mathrm{e}^{ \Psi(z)-\alpha_-(\phi(z)-\phi_\mathrm{b})}\right].
\end{split}
\label{eq:PB}
\end{equation}
Eq. \eqref{eq:bc} yields the boundary condition as given by 
\begin{equation}
-\varepsilon(\phi_0)\varepsilon_0 \left.\frac{d\psi}{dz}\right|_{z=0} = \sigma_0.
\label{eq:h}
\end{equation}
Eq.~\eqref{eq:chemsol} gives the conditions for the uniform chemical potential of the solvent, given as 
\begin{equation}
\begin{split}
&-\frac{(\varepsilon_2-\varepsilon_1)\varepsilon_0}{2k_\mathrm{B}T}\left(\frac{d\psi}{dz}\right)^2 +\alpha_+(c_+-c_\mathrm{b})+\alpha_-(c_--c_\mathrm{b})\\
&+\frac{1}{v_0}\left[\ln\frac{\phi}{1-\phi}-\ln\frac{\phi_\mathrm{b}}{1-\phi_\mathrm{b}}-2\chi(\phi-\phi_\mathrm{b})\right] \\
& -\frac{K}{k_\mathrm{B}T}\frac{d^2\phi}{dz^2}=0,
\end{split}
\label{eq:g}
\end{equation}
with the boundary condition 
\begin{equation}
\left.\frac{d\phi}{dz}\right|_{z=0}=0.
\label{eq:bcphi}
\end{equation}

To numerically obtain the equilibrium profiles for the case of $K\neq 0$, we solve the dynamic version of the equations \eqref{eq:PB} and \eqref{eq:g}, given by $\partial \psi/\partial t = \partial /\partial z(\varepsilon\varepsilon_0 \partial \psi/\partial z)+\cdots$ and $\partial \phi/\partial t = (K/k_\mathrm{B}T)\partial^2\phi/\partial z^2+\cdots$ with the help of the Crank-Nicolson method and the Thomas algorithm. 
We use a grid spacing of $\Delta z= 10^{-2}\,$nm.  
For the case $K=0$, we solve the pressure equation as given by
\begin{equation}
\begin{split}
&\frac{(\varepsilon_1+2(\varepsilon_2-\varepsilon_1)\phi)\varepsilon_0}{2k_\mathrm{B}T}\left(\frac{d\psi}{dz}\right)^2 -\frac{K}{k_\mathrm{B}T}\left[\frac{1}{2}\left(\frac{d\phi}{dz}\right)^2-\phi\frac{d^2\phi}{dz^2}\right]\\
&= c_++c_--2c_\mathrm{b}+\alpha_+(c_+\phi-c_\mathrm{b}\phi_\mathrm{b})+\alpha_-(c_-\phi-c_\mathrm{b}\phi_\mathrm{b}) \\
&-\frac{1}{v_0}\ln\frac{1-\phi}{1-\phi_\mathrm{b}}-\frac{1}{v_0}\chi(\phi^2-{\phi_\mathrm{b}}^2),
\end{split}
\label{eq:f}
\end{equation}
instead of the Poisson-Boltzmann equation. 
Eq.~\eqref{eq:f} can be derived by the Euler-Lagrange formalism, see Ref.~\citenum{Ben_Yaakov_2009} for details.
In the case of $K=0$, eq.~\ref{eq:bcphi} is not necessary. 
Equating $(d\psi/dz)^2$ from Eqs.~\eqref{eq:g} and \eqref{eq:f} gives an algebraic relation for $\psi(z)$ and $\phi(z)$. 
When $\psi_0=\psi(0)$ is specified, $\phi_0=\phi(0)$ and $\sigma_0$ can be obtained by solving Eqs.~\eqref{eq:g}, \eqref{eq:f} and \eqref{eq:h}.
To obtain the profiles of $\phi(z)$ and $\psi(z)$, we calculate $\psi(\Delta z) = \psi_0 + (d\psi/dz)|_{z=0}\Delta z$, and then repeat this procedure by solving \eqref{eq:g}, \eqref{eq:f} and \eqref{eq:h} at each step.
We use a  spatial step size of $\Delta z = 10^{-3}\,$nm for the numerical calculation in the case of $K=0$.

\section{Results and Discussion}

\subsection{Single-Liquid Approximation}
In the context of polymers in a binary mixture, the solvent quality is often evaluated using the single-liquid approximation (SLA) \cite{Scott_1949,Uematsu_2012}.
In this approximation, the solvent composition of the two coexisting phases is assumed to be the same, and the $\chi$ parameter for the interaction between the polymer and the binary mixture is approximated as the weighted average of the individual $\chi$ values for the two solvents.  
Applying this approximation to the electric double layer of binary mixtures, the solvent composition in the diffuse layer is assumed to be identical to that of the bulk phase. 
Then, the Debye wave number and the capacitance are given by
\begin{eqnarray}
\kappa_0 & = & \sqrt{\frac{2e^2c_\mathrm{b}}{\varepsilon(\phi_\mathrm{b})\varepsilon_0 k_\mathrm{B}T}},\label{eq:SLA_kappa}\\
C_0 & = & \varepsilon(\phi_\mathrm{b})\varepsilon_0 \kappa_0.\label{eq:SLA_capacitance}
\end{eqnarray} 
Eqs.~\eqref{eq:SLA_kappa} and \eqref{eq:SLA_capacitance} are used as reference expressions for the screening length and the capacitance in a binary mixture in the following analysis. 

\subsection{Analytical Results without the Gradient Free Energy}

In this section, we calculate the capacitance at the PZC ($\sigma_0=0$) without the gradient free energy ($K=0$).
The trivial PZC corresponds to $\psi_0=0$ ($\Psi_0=0$), where the profiles are $\phi(z)=\phi_\mathrm{b}$ and $\psi(z)=0$. 
Another PZC may exist, but we do not consider it in this work.
The linearization of eqs.~\eqref{eq:PB} and \eqref{eq:g} yield
\begin{equation}
\frac{d^2}{dz^2}\left(
\begin{array}{c}
\delta\phi \\
\delta\Psi
\end{array}
\right) = \mathsf{L}\left(
\begin{array}{c}
\delta\phi \\
\delta\Psi
\end{array}
\right),
\label{eq:diff1}
\end{equation}
where $\mathsf{L}$ is the matrix defined by 
\begin{equation}
\mathsf{L} = \left(
\begin{array}{cc}
k^2/S_0 & -k^2(\alpha_+-\alpha_-)c_\mathrm{b}v_0 \\
{\kappa_0}^2 (\alpha_+-\alpha_-)/2  & {\kappa_0}^2  
\end{array}
\right),
\end{equation}
and $k=\sqrt{k_\mathrm{B}T/Kv_0}$, and $1/S_0={\phi_\mathrm{b}}^{-1}+(1-\phi_\mathrm{b})^{-1}-2\chi-({\alpha_+}^2+{\alpha_-}^2)c_\mathrm{b}v_0$.
When we set $K=0$, eq.~\eqref{eq:diff1} becomes
\begin{equation}
\left(
\begin{array}{c}
0 \\
(d^2/dz^2)\delta\Psi
\end{array}
\right) = \mathsf{L}\left(
\begin{array}{c}
\delta\phi \\
\delta\Psi
\end{array}
\right),
\label{eq:diff2}
\end{equation}
and thus, $\delta\phi$ and $\delta\psi$ have a relation
\begin{equation}
\delta\phi(z) = S_0(\alpha_+-\alpha_-)c_\mathrm{b}v_0 \delta\Psi(z).\label{eq:30}
\end{equation}
The linearised Poisson-Boltzmann equation becomes
\begin{equation}
\frac{d^2\delta\Psi}{dz^2} = {\kappa_0}^2\left(1+\frac{S_0(\alpha_+-\alpha_-)^2c_\mathrm{b}v_0}{2}\right)\delta\Psi.
\end{equation}
This leads the capacitance at the PZC as
\begin{equation}
C = \left(\frac{d\sigma_0}{d\psi_0}\right)_{\psi_0=0} = \varepsilon(\phi_\mathrm{b})\varepsilon_0\kappa, \label{eq:analytic_capacitance}
\end{equation}
where
\begin{equation}
\kappa = \kappa_0\sqrt{1+\frac{S_0(\alpha_+-\alpha_-)^2c_\mathrm{b}v_0}{2}} \label{eq:kappa}.
\end{equation}
Eq.~\eqref{eq:kappa} demonstrates that in a binary mixture with $K=0$, the screening length generally differs from the Debye length in the SLA approximation, Eq.~\eqref{eq:SLA_kappa}.

\begin{figure}
\includegraphics[width=8cm]{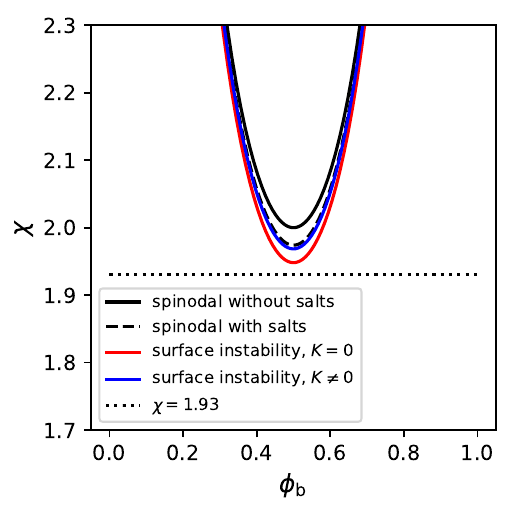}
\caption{
Phase diagram in the $\chi$–$\phi_\mathrm{b}$ plane.
The parameters are set as $\alpha_+ = 0$, $\alpha_- = 8$, $c_\mathrm{b} = 0.1\,$M, and $v_0 = 27\,\text{\AA}^3$.
The solid line indicates the spinodal line for a binary mixture without salt, corresponding to Eq.~\eqref{eq:bulk_instability} with $c_\mathrm{b} = 0$.
The dashed line represents the spinodal line for a binary mixture with salt, given by Eq.~\eqref{eq:bulk_instability}.
The red solid line denotes the surface instability line for $K = 0$ defined by Eq.~\eqref{eq:surface_instability}.
The blue solid line denotes the surface instability line for $K\neq 0$ defined by Eq.~\eqref{eq:instability2} with $K=k_\mathrm{B}T/{v_0}^{1/3}$.
The dotted line at $\chi = 1.93$ corresponds to the condition used in Fig.~\ref{fig:02}.
}
\label{fig:00}
\end{figure}

In the case of $\alpha_+=\alpha_-$, we have $\kappa=\kappa_0$. 
When $\alpha_+\neq\alpha_-$, the screening length deviates from $1/\kappa_0$.
Provided $S_0>0$, the screening length is shorter than $1/\kappa_0$. 
When the condition 
\begin{equation}
\frac{1}{S_0}=0,\label{eq:surface_instability}
\end{equation}
is satisfied, a interfacial instability arises, and the capacitance diverges. 
Note that the condition of eq.~\eqref{eq:surface_instability} is slightly different from that of the inverse structure factor at zero wavelength in previous works \cite{Onuki_2004,Onuki_2006,Onuki_2011},
\begin{equation}
{\phi_\mathrm{b}}^{-1}+(1-\phi_\mathrm{b})^{-1}-2\chi-\frac{(\alpha_++\alpha_-)^2}{2}c_\mathrm{b}v_0=0. \label{eq:bulk_instability}
\end{equation}
Because ${\alpha_+}^2+{\alpha_-}^2\ge (\alpha_++\alpha_-)^2/2$, an interfacial instability can occur even when the bulk phase remains stable.  
Fig.~\ref{fig:00} shows the phase diagram of the $\chi$-$\phi_\mathrm{b}$ plane. 
The parameters are set as $\alpha_+=0$, $\alpha_-=8$, $c_\mathrm{b}=0.1\,$M, and $v_0 =27\,$\AA$^3$.
The solid line indicates the spinodal line for binary mixtures without salts, corresponding to eq.~\eqref{eq:bulk_instability} with $c_\mathrm{b}=0$. 
The dashed line represents the spinodal line for binary mixture with salts, given by eq.~\eqref{eq:bulk_instability}.
The red solid line corresponds to the interfacial instability line for $K=0$ defined by eq.~\eqref{eq:surface_instability}.
These three lines are translations by a constant amount in the $\chi$ direction, and therefore never intersect.
The dotted line at $\chi=1.93$ corresponds to the condition used in Fig.~\ref{fig:02}.

\subsection{Analytical Results with the Gradient Free Energy}
When considering the gradient free energy, solving eq.~\eqref{eq:diff1} analytically is required. 
The eigenvalues of the matrix $\mathsf{L}$ are
\begin{equation}
\lambda_\pm = \frac{1}{2}\left(\frac{k^2}{S_0} +{\kappa_0}^2 \pm \sqrt{D}\right),
\end{equation}
where
\begin{equation}
D = \left(\frac{k^2}{S_0}-{\kappa_0}^2\right)^2-2{\kappa_0}^2k^2(\alpha_+-\alpha_-)^2 c_\mathrm{b}v_0.
\end{equation}
When  $\alpha_+=\alpha_-$, the matrix $\mathsf{L}$ is already diagonalized. 
Thus, $\phi(z)$ and $\Psi(z)$ are decoupled, and the solution is $\phi(z)=\phi_\mathrm{b}$ and $\Psi(z)=\Psi_0\mathrm{e}^{-\kappa_0 z}$.
When $\alpha_+\neq\alpha_-$, the eigenvectors are given by
\begin{equation}
\boldsymbol{\mathrm{v}}_\pm = \left(
\begin{array}{c}
k^2/S_0 -{\kappa_0}^2 \pm \sqrt{D} \\
{\kappa_0}^2(\alpha_+-\alpha_-)
\end{array}
\right),
\end{equation}
and the general solution for the differential equation, \eqref{eq:diff1} with $\alpha_+\neq \alpha_-$ is obtained as 
\begin{equation}
\left(
\begin{array}{c}
\delta\phi \\
\delta\Psi
\end{array}
\right) = C_+ \mathrm{e}^{-\sqrt{\lambda_+}z}\boldsymbol{\mathrm{v}}_+ + C_-\mathrm{e}^{-\sqrt{\lambda_-}z}\boldsymbol{\mathrm{v}}_-.\label{eq:kappa2}
\end{equation}
Thus, we have two screening lengths $\sqrt{\lambda_+}$ and $\sqrt{\lambda_-}$, and when they are complex conjugates, damped oscillation occurs.
The conditions of $\delta\phi'(0)=0$ and $\delta\Psi(0)=\Psi_0$ determine the constants $C_\pm$, as given by
\begin{eqnarray}
\frac{C_+}{ C_-} & = & -\frac{\sqrt{\lambda_-}\left(k^2/S_0 -{\kappa_0}^2 - \sqrt{D} \right)}{\sqrt{\lambda_+}\left(k^2/S_0 -{\kappa_0}^2 +\sqrt{D} \right)},\\
C_- & = & \frac{\Psi_0}{{\kappa_0}^2(\alpha_+-\alpha_-)}\left( 1+\frac{C_+}{C_-}\right)^{-1}.
\end{eqnarray}
The capacitance at the PZC is given by
\begin{equation}
C = \frac{2\varepsilon(\phi_\mathrm{b})\varepsilon_0\sqrt{\lambda_+\lambda_-D}}{\sqrt{\lambda_+}\left(\frac{k^2}{S_0} -{\kappa_0}^2 +\sqrt{D} \right)-\sqrt{\lambda_-}\left(\frac{k^2}{S_0} -{\kappa_0}^2 -\sqrt{D} \right)}. \label{eq:cap2}
\end{equation}
The condition for the denominator of eq.~\eqref{eq:cap2} to vanish  yields
\begin{equation}
\frac{1}{S_0} = \frac{{\kappa_0}^2-\sqrt{{\kappa_0}^4+2{\kappa_0}^2k^2(\alpha_+-\alpha_-)^2 c_\mathrm{b}v_0}}{2k^2},
\label{eq:instability2}
\end{equation} 
which is plotted by the blue solid line in Fig.~\ref{fig:00}, where we use $K=k_\mathrm{B}T/{v_0}^{1/3}$ and other parameter is the same as the red solid line.
The surface instability line for $K\neq 0$ differs from the case of $K=0$ (the dashed line), and the capacitance diverges when it approaches the surface instability line. 
In the limit of $2(k/\kappa_0)^2(\alpha_+-\alpha_-)^2c_\mathrm{b}v_0\ll 1$, eq.~\eqref{eq:instability2} is equivalent to eq.~\eqref{eq:bulk_instability}. 

\subsection{Numerical Results}

\begin{figure}
\includegraphics[width=8cm]{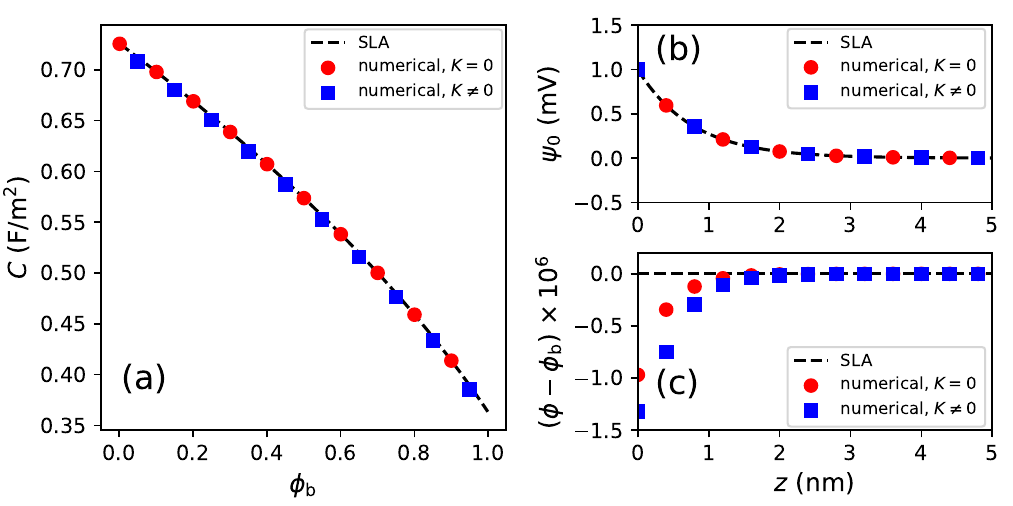}
\caption{
(a) Diffuse-layer capacitance as a function of the bulk solvent composition for the case $\alpha_+=\alpha_-=0$.
The other parameters are $c_\mathrm{b}=0.1\,$M, $\chi=0.5$, $\varepsilon_1=80$, $\varepsilon_2 = 20$, $v_0 = 27\,$\AA$^3$, $T=300\,$K, and $K=k_\mathrm{B}T/{v_0}^{1/3}$. 
The red and blue points are obtained numerically, whereas the dashed line is eq.~\eqref{eq:SLA_capacitance}.
(b) Electrostatic potential profile as a function of $z$ with the bulk composition is $\phi_\mathrm{b}=0.5$.
The points are obtained numerically, whereas the dashed line is eq.~\eqref{eq:debyehuckel} with $\kappa_0$.
(c) Solvent composition profile as a function of $z$ with the bulk compostion $\phi_\mathrm{b}=0.5$.
The points are obtained numerically, whereas the dashed line is the SLA result, $\phi=\phi_\mathrm{b}$. 
}
\label{fig:01}
\end{figure}

Fig.~\ref{fig:01}a shows the diffuse-layer capacitance as a function of the bulk solvent composition $\phi_\mathrm{b}$. 
The parameters used are $\alpha_+=\alpha_-=0$, $c_\mathrm{b}=0.1\,$M, $\chi=0.5$, $\varepsilon_1=80$, $\varepsilon_2 = 20$, $v_0 = 27\,$\AA$^3$, $T=300\,$K, and $K=k_\mathrm{B}T/{v_0}^{1/3}$. 
The points are obtained from the numerical calculation of eqs.~\eqref{eq:g} and \eqref{eq:PB} or \eqref{eq:f}.
The numerical capacitance is defined by $C = \sigma_0/\psi_0$, where $\psi_0 = 1\,$mV ($\Psi_0=0.039$).
The dashed line shows the analytical result using SLA, eq.~\eqref{eq:SLA_capacitance}.
The numerical results of both $K=0$ (red points) and $K\neq 0$ (blue points) agree with SLA because eq.~\eqref{eq:kappa} equals $\kappa_0$ when $\alpha_+=\alpha_-$. 
Fig.~\ref{fig:01}b and c show the profiles of $\psi(z)$ and $\phi(z)-\phi_\mathrm{b}$ at $\phi_\mathrm{b}=0.5$, and other parameters are the same as in Fig.~\ref{fig:01}a.  
The points represent the numerical results, while the dashed line is given by 
\begin{equation}
\psi(z)=\psi_0 \mathrm{e}^{-\kappa_0 z}. \label{eq:debyehuckel}
\end{equation}
The line agrees with the numerical results.
Thus, the bare Debye length, ${\kappa_0}^{-1}$, serves as the screening length in the binary mixture $\alpha_+=\alpha_-$.
The numerical composition profiles, $\phi(z)-\phi_\mathrm{b}$ shows adsorption of the first (high-dielectric) solvent. \cite{Ben_Yaakov_2009} 
The magnitude of $\phi_0-\phi_\mathrm{b}$ is quite small and is not reproduced by the SLA or other analytical linear theory.
Therefore, the deviation is the second-order effect in terms of $\delta\psi$.
These results demonstrate that unless $\chi$ and $\phi_\mathrm{b}$ are close to the critical or spinodal condition of the binary mixture, this deviation remains negligible.  

\begin{figure}
\includegraphics[width=8cm]{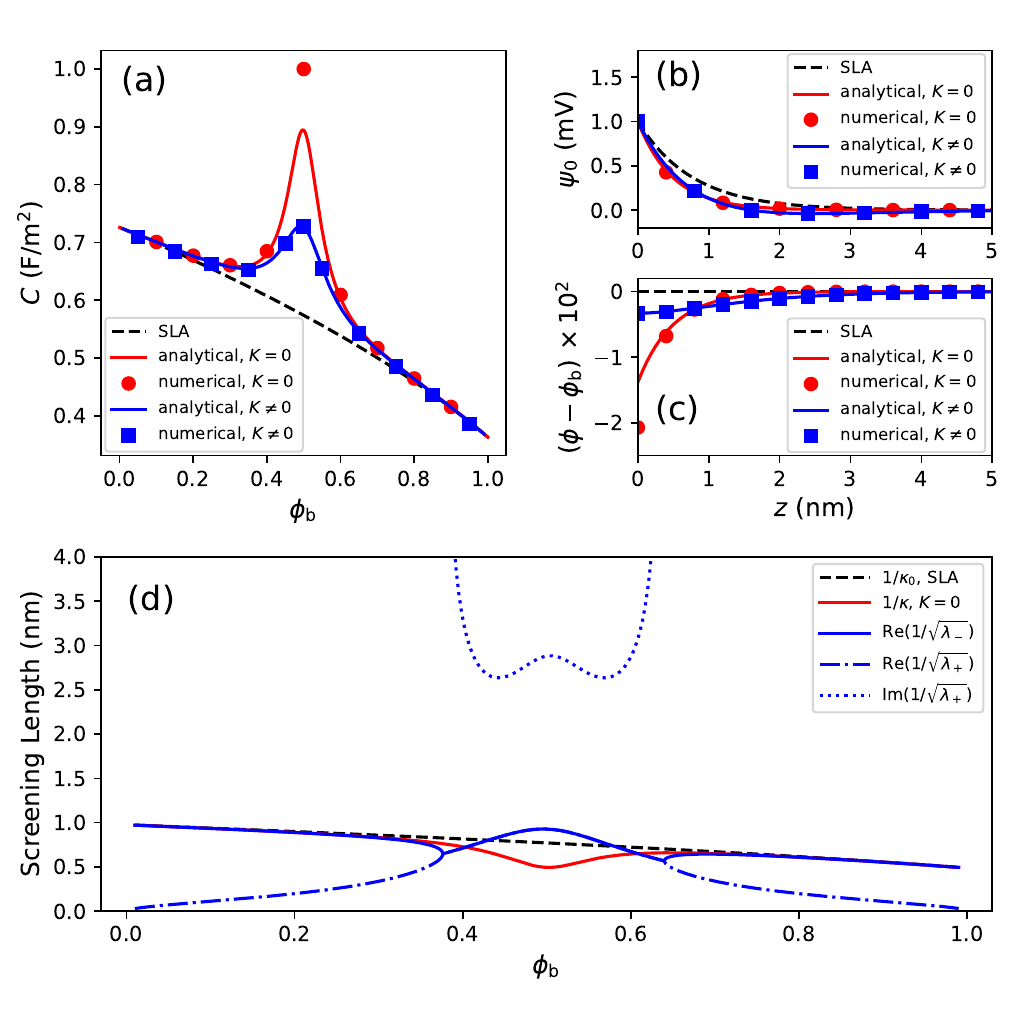}
\caption{
(a) Diffuse-layer capacitance as a function of the bulk solvent composition for the case $\alpha_+=0$ and $\alpha_-=8$.
The other parameters are $c_\mathrm{b}=0.1\,$M, $\chi=1.93$, $\varepsilon_1=80$, $\varepsilon_2 = 20$, $v_0 = 27\,$\AA$^3$, $T=300\,$K, and $K=k_\mathrm{B}T/{v_0}^{1/3}$. 
The points are obtained numerically, whereas the dashed line is eq.~\eqref{eq:SLA_capacitance} and the red solid line is eqs.~\eqref{eq:analytic_capacitance} and \eqref{eq:kappa}.
(b) Profile of the electrostatic potential as a function of $z$. The bulk composition is $\phi_\mathrm{b}=0.5$.
The points are obtained numerically, whereas the dased line is eq.~\eqref{eq:debyehuckel}  with $\kappa_0$, the red solid line is one with $\kappa$ in eq.~\eqref{eq:kappa}, and the blue solid line is eq.~\eqref{eq:kappa2}.
(c) Profile of the solvent composition as a function of $z$ with the bulk composition $\phi_\mathrm{b}=0.5$.
The points are obtained numerically, whereas the red solid line is calculated by eq.~\eqref{eq:30} and the blue solid line is eq.~\ref{eq:kappa2}.
(d) Volume fraction dependence of various screening lengths. The dashed black line is the result of SLA, the red solid line is the result of $K=0$, eq.~\ref{eq:kappa}, the blue solid and dashed-dotted lines are $\mathsf{Re}(1/\sqrt{\lambda_\pm})$, and the blue dotted line is $\mathsf{Im}(1/\sqrt{\lambda_+})$.
}
\label{fig:02}
\end{figure}

Fig.~\ref{fig:02}a shows the diffuse-layer capacitance when $\alpha_+=0$, $\alpha_-=8$, and $\chi=1.93$. 
Other parameters are the same as those in Fig.~\ref{fig:01}a.   
The points are obtained numerically as those in Fig.~\ref{fig:01}a, whereas the dashed black line is SLA capacitance, eq.~\eqref{eq:SLA_capacitance}.
When the bulk composition approaches $\phi_\mathrm{b}=0.5$, the numerical capacitances of both $K=0$ (red points) and $K\neq 0$ (blue points) deviate significantly from the SLA prediction. 
Furthermore, the analytic capacitances (the solid red and blue lines) are calculated using by Eqs.~\eqref{eq:analytic_capacitance} or \eqref{eq:cap2}, both of which show good agreement with the numerical results, respectively.
These results demonstrate that, when $\alpha_+\neq\alpha_-$, the inhomogeneity of the solvent composition near the surface affects the capacitance.
Fig.~\ref{fig:02}b and c show the profiles of $\psi(z)$ and $\phi(z)$ at $\phi_\mathrm{b}=0.5$.
The points are numerically obtained as in Fig.~\ref{fig:01}, where the black dashed, red solid, and blue solid lines are eqs.~\ref{eq:debyehuckel}, \eqref{eq:kappa} and \eqref{eq:kappa2}.
The Debye length $\kappa^{-1}$ of Eq.~\eqref{eq:kappa} is shorter than ${\kappa_0}^{-1}$, and $\kappa^{-1}$ actually works as the screening length of the double layer for the case $K=0$. 
In Fig.~\ref{fig:02}d, various screening lengths are plotted as a function of $\phi_\mathrm{b}$. 
The black dashed line shows the result of SLA, eq.~\ref{eq:SLA_kappa}, whereas the red solid line shows the result of $K=0$, eq.~\ref{eq:kappa}. 
When the system approaches the surface instability, the screening length $1/\kappa$ is smaller than $1/\kappa_0$. 
In the case of $K\neq 0$, the quantities $1/\sqrt{\lambda_\pm}$ may be complex. 
The real and imaginary parts of $1/\sqrt{\lambda_\pm}$ are represented by the blue solid, dashed-dotted, and dotted lines in Fig.~\ref{fig:02}d. 
When the system is close to the surface instability, $\mathsf{Re}(1/\sqrt{\lambda_-})$ is larger than $1/\kappa_0$. 
At that time, the wavelength of the damped oscillation $\mathsf{Im}(1/\sqrt{\lambda_+})$ is approximately three times as large as $\mathsf{Re}(1/\sqrt{\lambda_-})$. 
As a result, the profile of $\psi(z)$ (blue line in b) shows an effective single screening length smaller than $1/\kappa_0$.  
The profiles of $\phi$ in both cases of $K=0$ and $K\neq 0$ also show adsorption of the first solvent, and the gradient term suppresses the steep variation in $\phi(z)$.

\section{Conclusion}

In this work, we analytically obtain the diffuse-layer capacitance at the PZC in binary mixtures.
Furthermore, we find a divergence in the capacitance on the phase diagram, suggesting a surface instability at the PZC. 
The obtained capacitance differs from that derived using the SLA unless the preferential solvation effect is symmetric for cations and anions.  
When $\alpha_+\neq\alpha_-$ and $\chi$ is close to the critical point, the capacitance at PZC significantly deviates from the SLA prediction, reflecting the surface instability we identified.

Here, we make some remarks and outline future directions. 
First, Yabunaka, et al. reported another type of surface instability and a phase transition even at $\chi<0$ when the salt is antagonistic ($\alpha_+=10$ and $\alpha_-=-10$) \cite{Yabunaka_2017}.
Because our analysis focuses solely on the capacitance at the PZC, the relationship between the instability we identified and that reported in Ref.~\citenum{Yabunaka_2017} remains unclear.
We will investigate the differential capacitance of binary mixtures in the future to clarify this relationship. 

Second, our analysis demonstrates that the presence of the gradient free energy slightly modifies the quantitative results, but the qualitative behavior remains unchanged.  
Typically, the gradient free energy has been considered for near-critical binary mixtures, such as water-3-methylpyridine mixture \cite{Sadakane_2009} at a temperature and composition near the critical point.
However, a water-ethanol mixture at room temperature is far from criticality for any composition.
Quantitative evaluation of $K$ for such binary mixtures far from criticality  is challenging.  

Third, the experimental results on the capacitance in binary mixtures exhibit more distinct behavior that cannot be explained by the SLA. 
Although we also observe significant deviation from the SLA prediction, as shown in Fig.~\ref{fig:02}, it is caused by surface instability of the binary mixture.
However, in the experimental data, the capacitance as a function of the composition exhibits distinct variation only in the range of $\phi_\mathrm{b}=0.8$ to $1$.
For other volume fractions, the capacitance is similar to that in pure water \cite{Aoki_2018, Iwasaki_2023}. 
Thus, our theory is insufficient to explain the experimental trend. 
Further investigation is needed, and the discrepancy might be resolved by considering the Stern-layer capacitance in binary mixtures. 

We hope that our analytical theory and discovery of the divergence of the capacitance in binary mixtures help the development of supercapacitors and other double-layer-related devices.

\begin{acknowledgments}
The author thanks financial support from JSPS KAKENHI (Grant No. 23K13073).
\end{acknowledgments}

\section*{DATA AVAILABILITY}
The data that support the findings of this study are available from the corresponding author upon reasonable request.

\bibliography{electrochem}

\begin{thebibliography}{19}%
\makeatletter
\providecommand \@ifxundefined [1]{%
 \@ifx{#1\undefined}
}%
\providecommand \@ifnum [1]{%
 \ifnum #1\expandafter \@firstoftwo
 \else \expandafter \@secondoftwo
 \fi
}%
\providecommand \@ifx [1]{%
 \ifx #1\expandafter \@firstoftwo
 \else \expandafter \@secondoftwo
 \fi
}%
\providecommand \natexlab [1]{#1}%
\providecommand \enquote  [1]{``#1''}%
\providecommand \bibnamefont  [1]{#1}%
\providecommand \bibfnamefont [1]{#1}%
\providecommand \citenamefont [1]{#1}%
\providecommand \href@noop [0]{\@secondoftwo}%
\providecommand \href [0]{\begingroup \@sanitize@url \@href}%
\providecommand \@href[1]{\@@startlink{#1}\@@href}%
\providecommand \@@href[1]{\endgroup#1\@@endlink}%
\providecommand \@sanitize@url [0]{\catcode `\\12\catcode `\$12\catcode
  `\&12\catcode `\#12\catcode `\^12\catcode `\_12\catcode `\%12\relax}%
\providecommand \@@startlink[1]{}%
\providecommand \@@endlink[0]{}%
\providecommand \url  [0]{\begingroup\@sanitize@url \@url }%
\providecommand \@url [1]{\endgroup\@href {#1}{\urlprefix }}%
\providecommand \urlprefix  [0]{URL }%
\providecommand \Eprint [0]{\href }%
\providecommand \doibase [0]{https://doi.org/}%
\providecommand \selectlanguage [0]{\@gobble}%
\providecommand \bibinfo  [0]{\@secondoftwo}%
\providecommand \bibfield  [0]{\@secondoftwo}%
\providecommand \translation [1]{[#1]}%
\providecommand \BibitemOpen [0]{}%
\providecommand \bibitemStop [0]{}%
\providecommand \bibitemNoStop [0]{.\EOS\space}%
\providecommand \EOS [0]{\spacefactor3000\relax}%
\providecommand \BibitemShut  [1]{\csname bibitem#1\endcsname}%
\let\auto@bib@innerbib\@empty
\bibitem [{\citenamefont {Uematsu}, \citenamefont {Netz},\ and\ \citenamefont
  {Bonthuis}(2018)}]{Uematsu_2018_1}%
  \BibitemOpen
  \bibfield  {author} {\bibinfo {author} {\bibfnamefont {Y.}~\bibnamefont
  {Uematsu}}, \bibinfo {author} {\bibfnamefont {R.~R.}\ \bibnamefont {Netz}},\
  and\ \bibinfo {author} {\bibfnamefont {D.~J.}\ \bibnamefont {Bonthuis}},\
  }\bibfield  {title} {\enquote {\bibinfo {title} {Analytical interfacial layer
  model for the capacitance and electrokinetics of charged aqueous
  interfaces},}\ }\href {https://doi.org/10.1021/acs.langmuir.7b04171}
  {\bibfield  {journal} {\bibinfo  {journal} {Langmuir}\ }\textbf {\bibinfo
  {volume} {34}},\ \bibinfo {pages} {9097--9113} (\bibinfo {year}
  {2018})}\BibitemShut {NoStop}%
\bibitem [{\citenamefont {Uematsu}(2021)}]{Uematsu_2021}%
  \BibitemOpen
  \bibfield  {author} {\bibinfo {author} {\bibfnamefont {Y.}~\bibnamefont
  {Uematsu}},\ }\bibfield  {title} {\enquote {\bibinfo {title} {Electrification
  of water interface},}\ }\href {https://doi.org/10.1088/1361-648x/ac15d5}
  {\bibfield  {journal} {\bibinfo  {journal} {J. Phys.: Condens. Matter}\
  }\textbf {\bibinfo {volume} {33}},\ \bibinfo {pages} {423001} (\bibinfo
  {year} {2021})}\BibitemShut {NoStop}%
\bibitem [{\citenamefont {Becker}\ \emph {et~al.}(2023)\citenamefont {Becker},
  \citenamefont {Loche}, \citenamefont {Rezaei}, \citenamefont {Wolde-Kidan},
  \citenamefont {Uematsu}, \citenamefont {Netz},\ and\ \citenamefont
  {Bonthuis}}]{Becker_2023}%
  \BibitemOpen
  \bibfield  {author} {\bibinfo {author} {\bibfnamefont {M.}~\bibnamefont
  {Becker}}, \bibinfo {author} {\bibfnamefont {P.}~\bibnamefont {Loche}},
  \bibinfo {author} {\bibfnamefont {M.}~\bibnamefont {Rezaei}}, \bibinfo
  {author} {\bibfnamefont {A.}~\bibnamefont {Wolde-Kidan}}, \bibinfo {author}
  {\bibfnamefont {Y.}~\bibnamefont {Uematsu}}, \bibinfo {author} {\bibfnamefont
  {R.~R.}\ \bibnamefont {Netz}},\ and\ \bibinfo {author} {\bibfnamefont
  {D.~J.}\ \bibnamefont {Bonthuis}},\ }\bibfield  {title} {\enquote {\bibinfo
  {title} {Multiscale modeling of aqueous electric double layers},}\ }\href
  {https://doi.org/10.1021/acs.chemrev.3c00307} {\bibfield  {journal} {\bibinfo
   {journal} {Chemical Reviews}\ }\textbf {\bibinfo {volume} {124}},\ \bibinfo
  {pages} {1–26} (\bibinfo {year} {2023})}\BibitemShut {NoStop}%
\bibitem [{\citenamefont {Gouy}(1910)}]{Gouy_1910}%
  \BibitemOpen
  \bibfield  {author} {\bibinfo {author} {\bibfnamefont {M.}~\bibnamefont
  {Gouy}},\ }\bibfield  {title} {\enquote {\bibinfo {title} {Sur la
  constitution de la charge {\'{e}}lectrique {\`{a}} la surface
  d{\textquotesingle}un {\'{e}}lectrolyte},}\ }\href
  {https://doi.org/10.1051/jphystap:019100090045700} {\bibfield  {journal}
  {\bibinfo  {journal} {J. Phys. Theor. Appl.}\ }\textbf {\bibinfo {volume}
  {9}},\ \bibinfo {pages} {457--468} (\bibinfo {year} {1910})}\BibitemShut
  {NoStop}%
\bibitem [{\citenamefont {Chapman}(1913)}]{Chapman_1913}%
  \BibitemOpen
  \bibfield  {author} {\bibinfo {author} {\bibfnamefont {D.~L.}\ \bibnamefont
  {Chapman}},\ }\bibfield  {title} {\enquote {\bibinfo {title} {A contribution
  to the theory of electrocapillarity},}\ }\href
  {https://doi.org/10.1080/14786440408634187} {\bibfield  {journal} {\bibinfo
  {journal} {The London, Edinburgh, and Dublin Philosophical Magazine and
  Journal of Science}\ }\textbf {\bibinfo {volume} {25}},\ \bibinfo {pages}
  {475--481} (\bibinfo {year} {1913})}\BibitemShut {NoStop}%
\bibitem [{\citenamefont {Grahame}(1947)}]{Grahame_1947}%
  \BibitemOpen
  \bibfield  {author} {\bibinfo {author} {\bibfnamefont {D.~C.}\ \bibnamefont
  {Grahame}},\ }\bibfield  {title} {\enquote {\bibinfo {title} {The electrical
  double layer and the theory of electrocapillarity.}}\ }\href
  {https://doi.org/10.1021/cr60130a002} {\bibfield  {journal} {\bibinfo
  {journal} {Chem. Rev.}\ }\textbf {\bibinfo {volume} {41}},\ \bibinfo {pages}
  {441--501} (\bibinfo {year} {1947})}\BibitemShut {NoStop}%
\bibitem [{\citenamefont {Onuki}\ and\ \citenamefont
  {Kitamura}(2004)}]{Onuki_2004}%
  \BibitemOpen
  \bibfield  {author} {\bibinfo {author} {\bibfnamefont {A.}~\bibnamefont
  {Onuki}}\ and\ \bibinfo {author} {\bibfnamefont {H.}~\bibnamefont
  {Kitamura}},\ }\bibfield  {title} {\enquote {\bibinfo {title} {Solvation
  effects in near-critical binary mixtures},}\ }\href
  {https://doi.org/10.1063/1.1769357} {\bibfield  {journal} {\bibinfo
  {journal} {The Journal of Chemical Physics}\ }\textbf {\bibinfo {volume}
  {121}},\ \bibinfo {pages} {3143–3151} (\bibinfo {year} {2004})}\BibitemShut
  {NoStop}%
\bibitem [{\citenamefont {Onuki}(2006)}]{Onuki_2006}%
  \BibitemOpen
  \bibfield  {author} {\bibinfo {author} {\bibfnamefont {A.}~\bibnamefont
  {Onuki}},\ }\bibfield  {title} {\enquote {\bibinfo {title} {Ginzburg-landau
  theory of solvation in polar fluids: Ion distribution around an interface},}\
  }\href {https://doi.org/10.1103/physreve.73.021506} {\bibfield  {journal}
  {\bibinfo  {journal} {Physical Review E}\ }\textbf {\bibinfo {volume} {73}}
  (\bibinfo {year} {2006}),\ 10.1103/physreve.73.021506}\BibitemShut {NoStop}%
\bibitem [{\citenamefont {Ben-Yaakov}\ \emph {et~al.}(2009)\citenamefont
  {Ben-Yaakov}, \citenamefont {Andelman}, \citenamefont {Harries},\ and\
  \citenamefont {Podgornik}}]{Ben_Yaakov_2009}%
  \BibitemOpen
  \bibfield  {author} {\bibinfo {author} {\bibfnamefont {D.}~\bibnamefont
  {Ben-Yaakov}}, \bibinfo {author} {\bibfnamefont {D.}~\bibnamefont
  {Andelman}}, \bibinfo {author} {\bibfnamefont {D.}~\bibnamefont {Harries}},\
  and\ \bibinfo {author} {\bibfnamefont {R.}~\bibnamefont {Podgornik}},\
  }\bibfield  {title} {\enquote {\bibinfo {title} {Ions in mixed dielectric
  solvents: Density profiles and osmotic pressure between charged
  interfaces},}\ }\href {https://doi.org/10.1021/jp9003533} {\bibfield
  {journal} {\bibinfo  {journal} {The Journal of Physical Chemistry B}\
  }\textbf {\bibinfo {volume} {113}},\ \bibinfo {pages} {6001–6011} (\bibinfo
  {year} {2009})}\BibitemShut {NoStop}%
\bibitem [{\citenamefont {Onuki}, \citenamefont {Araki},\ and\ \citenamefont
  {Okamoto}(2011)}]{Onuki_2011}%
  \BibitemOpen
  \bibfield  {author} {\bibinfo {author} {\bibfnamefont {A.}~\bibnamefont
  {Onuki}}, \bibinfo {author} {\bibfnamefont {T.}~\bibnamefont {Araki}},\ and\
  \bibinfo {author} {\bibfnamefont {R.}~\bibnamefont {Okamoto}},\ }\bibfield
  {title} {\enquote {\bibinfo {title} {Solvation effects in phase transitions
  in soft matter},}\ }\href {https://doi.org/10.1088/0953-8984/23/28/284113}
  {\bibfield  {journal} {\bibinfo  {journal} {Journal of Physics: Condensed
  Matter}\ }\textbf {\bibinfo {volume} {23}},\ \bibinfo {pages} {284113}
  (\bibinfo {year} {2011})}\BibitemShut {NoStop}%
\bibitem [{\citenamefont {Aoki}, \citenamefont {Chen},\ and\ \citenamefont
  {Tang}(2018)}]{Aoki_2018}%
  \BibitemOpen
  \bibfield  {author} {\bibinfo {author} {\bibfnamefont {K.~J.}\ \bibnamefont
  {Aoki}}, \bibinfo {author} {\bibfnamefont {J.}~\bibnamefont {Chen}},\ and\
  \bibinfo {author} {\bibfnamefont {P.}~\bibnamefont {Tang}},\ }\bibfield
  {title} {\enquote {\bibinfo {title} {Double layer impedance in mixtures of
  acetonitrile and water},}\ }\href {https://doi.org/10.1002/elan.201800025}
  {\bibfield  {journal} {\bibinfo  {journal} {Electroanalysis}\ }\textbf
  {\bibinfo {volume} {30}},\ \bibinfo {pages} {1634--1641} (\bibinfo {year}
  {2018})}\BibitemShut {NoStop}%
\bibitem [{\citenamefont {Iwasaki}, \citenamefont {Kimura},\ and\ \citenamefont
  {Uematsu}(2023)}]{Iwasaki_2023}%
  \BibitemOpen
  \bibfield  {author} {\bibinfo {author} {\bibfnamefont {H.}~\bibnamefont
  {Iwasaki}}, \bibinfo {author} {\bibfnamefont {Y.}~\bibnamefont {Kimura}},\
  and\ \bibinfo {author} {\bibfnamefont {Y.}~\bibnamefont {Uematsu}},\
  }\bibfield  {title} {\enquote {\bibinfo {title} {Ubiquitous preferential
  water adsorption to electrodes in water/1-propanol mixtures detected by
  electrochemical impedance spectroscopy},}\ }\href
  {https://doi.org/10.1021/acs.jpcc.3c05320} {\bibfield  {journal} {\bibinfo
  {journal} {The Journal of Physical Chemistry C}\ }\textbf {\bibinfo {volume}
  {127}},\ \bibinfo {pages} {23382–23389} (\bibinfo {year}
  {2023})}\BibitemShut {NoStop}%
\bibitem [{\citenamefont {Pousaneh}, \citenamefont {Ciach},\ and\ \citenamefont
  {Maciołek}(2012)}]{Pousaneh_2012}%
  \BibitemOpen
  \bibfield  {author} {\bibinfo {author} {\bibfnamefont {F.}~\bibnamefont
  {Pousaneh}}, \bibinfo {author} {\bibfnamefont {A.}~\bibnamefont {Ciach}},\
  and\ \bibinfo {author} {\bibfnamefont {A.}~\bibnamefont {Maciołek}},\
  }\bibfield  {title} {\enquote {\bibinfo {title} {Effect of ions on confined
  near-critical binary aqueous mixture},}\ }\href
  {https://doi.org/10.1039/c2sm25461a} {\bibfield  {journal} {\bibinfo
  {journal} {Soft Matter}\ }\textbf {\bibinfo {volume} {8}},\ \bibinfo {pages}
  {7567} (\bibinfo {year} {2012})}\BibitemShut {NoStop}%
\bibitem [{\citenamefont {Pousaneh}, \citenamefont {Ciach},\ and\ \citenamefont
  {Maciołek}(2014)}]{Pousaneh_2014}%
  \BibitemOpen
  \bibfield  {author} {\bibinfo {author} {\bibfnamefont {F.}~\bibnamefont
  {Pousaneh}}, \bibinfo {author} {\bibfnamefont {A.}~\bibnamefont {Ciach}},\
  and\ \bibinfo {author} {\bibfnamefont {A.}~\bibnamefont {Maciołek}},\
  }\bibfield  {title} {\enquote {\bibinfo {title} {How ions in solution can
  change the sign of the critical casimir potential},}\ }\href
  {https://doi.org/10.1039/c3sm51972d} {\bibfield  {journal} {\bibinfo
  {journal} {Soft Matter}\ }\textbf {\bibinfo {volume} {10}},\ \bibinfo {pages}
  {470–483} (\bibinfo {year} {2014})}\BibitemShut {NoStop}%
\bibitem [{\citenamefont {Yabunaka}\ and\ \citenamefont
  {Onuki}(2017)}]{Yabunaka_2017}%
  \BibitemOpen
  \bibfield  {author} {\bibinfo {author} {\bibfnamefont {S.}~\bibnamefont
  {Yabunaka}}\ and\ \bibinfo {author} {\bibfnamefont {A.}~\bibnamefont
  {Onuki}},\ }\bibfield  {title} {\enquote {\bibinfo {title} {Electric double
  layer composed of an antagonistic salt in an aqueous mixture: Local charge
  separation and surface phase transition},}\ }\href
  {https://doi.org/10.1103/physrevlett.119.118001} {\bibfield  {journal}
  {\bibinfo  {journal} {Phys. Rev. Lett.}\ }\textbf {\bibinfo {volume} {119}},\
  \bibinfo {pages} {118001} (\bibinfo {year} {2017})}\BibitemShut {NoStop}%
\bibitem [{\citenamefont {Abrashkin}, \citenamefont {Andelman},\ and\
  \citenamefont {Orland}(2007)}]{Abrashkin_2007}%
  \BibitemOpen
  \bibfield  {author} {\bibinfo {author} {\bibfnamefont {A.}~\bibnamefont
  {Abrashkin}}, \bibinfo {author} {\bibfnamefont {D.}~\bibnamefont
  {Andelman}},\ and\ \bibinfo {author} {\bibfnamefont {H.}~\bibnamefont
  {Orland}},\ }\bibfield  {title} {\enquote {\bibinfo {title} {Dipolar
  poisson-boltzmann equation: Ions and dipoles close to charge interfaces},}\
  }\href {https://doi.org/10.1103/physrevlett.99.077801} {\bibfield  {journal}
  {\bibinfo  {journal} {Physical Review Letters}\ }\textbf {\bibinfo {volume}
  {99}} (\bibinfo {year} {2007}),\ 10.1103/physrevlett.99.077801}\BibitemShut
  {NoStop}%
\bibitem [{\citenamefont {Scott}(1949)}]{Scott_1949}%
  \BibitemOpen
  \bibfield  {author} {\bibinfo {author} {\bibfnamefont {R.~L.}\ \bibnamefont
  {Scott}},\ }\bibfield  {title} {\enquote {\bibinfo {title} {The
  thermodynamics of high polymer solutions. iv. phase equilibria in the ternary
  system: Polymer—liquid 1—liquid 2},}\ }\href
  {https://doi.org/10.1063/1.1747238} {\bibfield  {journal} {\bibinfo
  {journal} {The Journal of Chemical Physics}\ }\textbf {\bibinfo {volume}
  {17}},\ \bibinfo {pages} {268–279} (\bibinfo {year} {1949})}\BibitemShut
  {NoStop}%
\bibitem [{\citenamefont {Uematsu}\ and\ \citenamefont
  {Araki}(2012)}]{Uematsu_2012}%
  \BibitemOpen
  \bibfield  {author} {\bibinfo {author} {\bibfnamefont {Y.}~\bibnamefont
  {Uematsu}}\ and\ \bibinfo {author} {\bibfnamefont {T.}~\bibnamefont
  {Araki}},\ }\bibfield  {title} {\enquote {\bibinfo {title} {Effects of
  strongly selective additives on volume phase transition in gels},}\ }\href
  {https://doi.org/10.1063/1.4732857} {\bibfield  {journal} {\bibinfo
  {journal} {The Journal of Chemical Physics}\ }\textbf {\bibinfo {volume}
  {137}} (\bibinfo {year} {2012}),\ 10.1063/1.4732857}\BibitemShut {NoStop}%
\bibitem [{\citenamefont {Sadakane}\ \emph {et~al.}(2009)\citenamefont
  {Sadakane}, \citenamefont {Onuki}, \citenamefont {Nishida}, \citenamefont
  {Koizumi},\ and\ \citenamefont {Seto}}]{Sadakane_2009}%
  \BibitemOpen
  \bibfield  {author} {\bibinfo {author} {\bibfnamefont {K.}~\bibnamefont
  {Sadakane}}, \bibinfo {author} {\bibfnamefont {A.}~\bibnamefont {Onuki}},
  \bibinfo {author} {\bibfnamefont {K.}~\bibnamefont {Nishida}}, \bibinfo
  {author} {\bibfnamefont {S.}~\bibnamefont {Koizumi}},\ and\ \bibinfo {author}
  {\bibfnamefont {H.}~\bibnamefont {Seto}},\ }\bibfield  {title} {\enquote
  {\bibinfo {title} {Multilamellar structures induced by hydrophilic and
  hydrophobic ions added to a binary mixture of d$_2$o and 3-methylpyridine},}\
  }\href {https://doi.org/10.1103/physrevlett.103.167803} {\bibfield  {journal}
  {\bibinfo  {journal} {Physical Review Letters}\ }\textbf {\bibinfo {volume}
  {103}} (\bibinfo {year} {2009}),\ 10.1103/physrevlett.103.167803}\BibitemShut
  {NoStop}%
\end{thebibliography}%
\end{document}